\documentclass[9pt]{article}
\usepackage{spconf,amsmath,graphicx}
\usepackage{multicol}
\usepackage{multirow}
\usepackage{makecell}
\usepackage{CJKutf8}
\usepackage{booktabs}
\usepackage{tabularx} % 自动宽度表格
\usepackage{ragged2e} % 文本对齐增强（可选）
\usepackage{booktabs} % \toprule \midrule \bottomrule \cmidrule{start-end} 表格的 start 到 end 列添加粗线
\usepackage{diagbox} % 创建具有斜线表头的表格单元格 \diagbox{Col 1}{Col 2}.
\title{Joint decoding method for controllable contextual speech recognition based on Speech LLM}
\name{Yangui Fang$^{1,4*}$\thanks{* Yangui Fang and Jing Peng contributed equally to this work.}, Jing Peng$^{2,3*}$, Yu Xi$^{2,3}$, Xu Li$^4$ ,Haoyu Li$^{2,3}$, Chengwei Zhang$^1$$^{\dagger}$,Guohui Zhong$^1$, Kai Yu$^{2,3}$\thanks {$^{\dagger}$ Corresponding Author.}}

\address{
\normalsize
$^1$School of Electronic Information and Communications, Huazhong University of Science and Technology, China\\
\normalsize
$^2$MoE Key Lab of Artificial Intelligence, AI Institute, X-LANCE Lab, Shanghai Jiao Tong University, Shanghai, China\\
\normalsize
$^3$Jiangsu Key Lab of Language Computing, Suzhou, China \quad $^4$AISpeech Ltd, Suzhou, China\\
\normalsize
\{fangyg,zhangcw,zhonggh\}@hust.edu.cn~~~danieljingpeng@gmail.com~~~\{yuxi.cs,haoyuli.cs,kai.yu\}@sjtu.edu.cn~~~ xu.li@aispeech.com~~~ 
}

\begin{document}
%\ninept
%
\maketitle

\begin{abstract}

% 语音偏执识别是指的是在带有上下文信息的情况下实现对特定内容的偏好识别。最近借助语音大语言模型的上下文理解能力，通过提示词（Prompt）的方式注入热词、领域知识等信息以实现语境偏向（contextual biasing）的语音识别，已成为研究热点且被证实有效。然而，直接通过提示词注入信息的方式依赖于Attention内部的注意力机制无法显式控制信息的注入程度。针对上述问题，我们提出了一种采用联合解码的热词信息注入识别解码方法，并且提供了多种策略，通过该方法我们可以控制注入的热词信息，同时获得更好的识别结果。并且实验证明对于未经过上下文训练的语音大模型也可以通过我们的方法获得上下文能力，同时我们开展了广泛的实验，系统性的研究了有关上下文的种种问题。
Contextual speech recognition refers to the ability to identify preferences for specific content based on contextual information. Recently, leveraging the contextual understanding capabilities of Speech LLM to  achieve contextual biasing by injecting contextual information through prompts have emerged as a research hotspot.
However, the direct information injection method via prompts relies on the internal attention mechanism of the model, making it impossible to explicitly control the extent of information injection. To address this limitation, we propose a joint decoding method to control the contextual information. This approach enables explicit control over the injected contextual information and achieving superior recognition performance. Additionally, Our method can also be used for sensitive word suppression recognition.
Furthermore, experimental results show that even Speech LLM not pre-trained on long contextual data can acquire long contextual capabilities through our method. 
% In recent years, the speech large model method, which connects the speech encoder with the large language model through a mapper, has achieved state-of-the-art (SOTA) results in the field of speech recognition. Meanwhile, speech recognition that achieves contextual biasing by injecting information such as hotwords and domain knowledge through Prompt with the help of the context understanding ability of large language models has become a research hotspot and been proven effective. However, the current simple method of directly using Prompt has two problems: (1) When the injected Prompt information is too long, the hallucination phenomenon may occur; (2) The way of injecting information through Prompt cannot explicitly control the degree of information injection. To address the above issues, we propose a hotword information injection recognition and decoding method using joint decoding. This method can suppress the hallucination problem by maintaining [there may be a typo here in the original text, presumably "a specific mechanism" or similar expressions]. Using hotword information as contextual information, we verified the effectiveness of the proposed method on the open-source SOTA Speech LLM and self-trained models based on the AISHELL-1/AISHELL-2 datasets.
\end{abstract}
\section{Introduction}

% 近年来，一种基于语音编码器、通过Adapter与大语言模型衔接的语音大语言模型方法，在语音识别任务中取得显著突破，实现了当前语音识别的SOTA性能\cite{chen2024salm,yu2024connecting,fathullah2024prompting,chu2024qwen2,xu2025fireredasr}。

% 语音识别领域正经历第三次范式转变：从传统语音识别系统\cite{levinson1983introduction,lee1989speaker,9536732}，到端到端语音识别框架\cite{abdel2014convolutional,gulati2020conformer,gao2022paraformer,graves2005framewise,ren2022improving}，再到语音大模型范式\cite{peng2025surveyspeechlargelanguage, arora2025landscapespokenlanguagemodels}。

% 在语音识别中，**语音偏置识别**是一个尤为关键的问题，其核心在于利用给定上下文信息提升特定词语（尤其是人名、地名等实体词）的识别精度\cite{aleksic2015bringing,hall2015composition,han2021cif,zhao2019shallow}。传统语音识别模型通常由语音模型与语言模型构成，Google团队\cite{aleksic2015bringing,hall2024composition}提出On-the-fly Rescore方法，通过在解码阶段利用WFST调整语言模型对特定词语的识别概率。进入深度学习时代后，语言模型虽不再是识别系统的必需组件，但仍能通过Shallow-Fusion等方法为系统提供上下文偏置\cite{zhao2019shallow}；另一种思路是深层融合，例如Google提出的CLAS方法\cite{pundak2018deep}，将特定领域的热词信息嵌入端到端模型的内部架构，实现语义信息与声学特征的深度交互。在大模型时代，得益于大语言模型卓越的上下文理解能力，在Prompt中嵌入上下文信息成为主流方式，多项研究已验证该方法的有效性\cite{chen2024salm,yang2024mala,shen2025retrieval,lakomkin2024end}。

% 然而，当前通过Prompt直接注入上下文信息的方式依赖隐式的Attention机制，难以精确控制热词信息的融合强度。在不同场景下，对热词信息的强调需求存在差异，且过度强调热词可能干扰整体识别精度。因此，一个关键问题应运而生：能否在语音大模型的基础上显式控制热词信息的融合程度，以实现更优的识别结果？甚至能否抑制不希望出现的敏感词？

% 为解决上述问题，本文提出联合解码方法，通过维护含上下文与不含上下文的两个解码束并设计合并策略，实现对热词融合程度的控制。该方法在AISHELL-1数据集上的性能优于单纯通过Prompt提示的方法；同时，在略微降低识别错误率的前提下，可有效抑制敏感词的识别。此外，本文系统性研究了合并策略与语音大模型上下文长度能力的相关问题。最后我们发现即使未经过上下文训练的语音大模型，也能通过该方法获得上下文偏置能力。 
\begin{figure*}[!ht]
\centering
\includegraphics[width=0.99\textwidth]{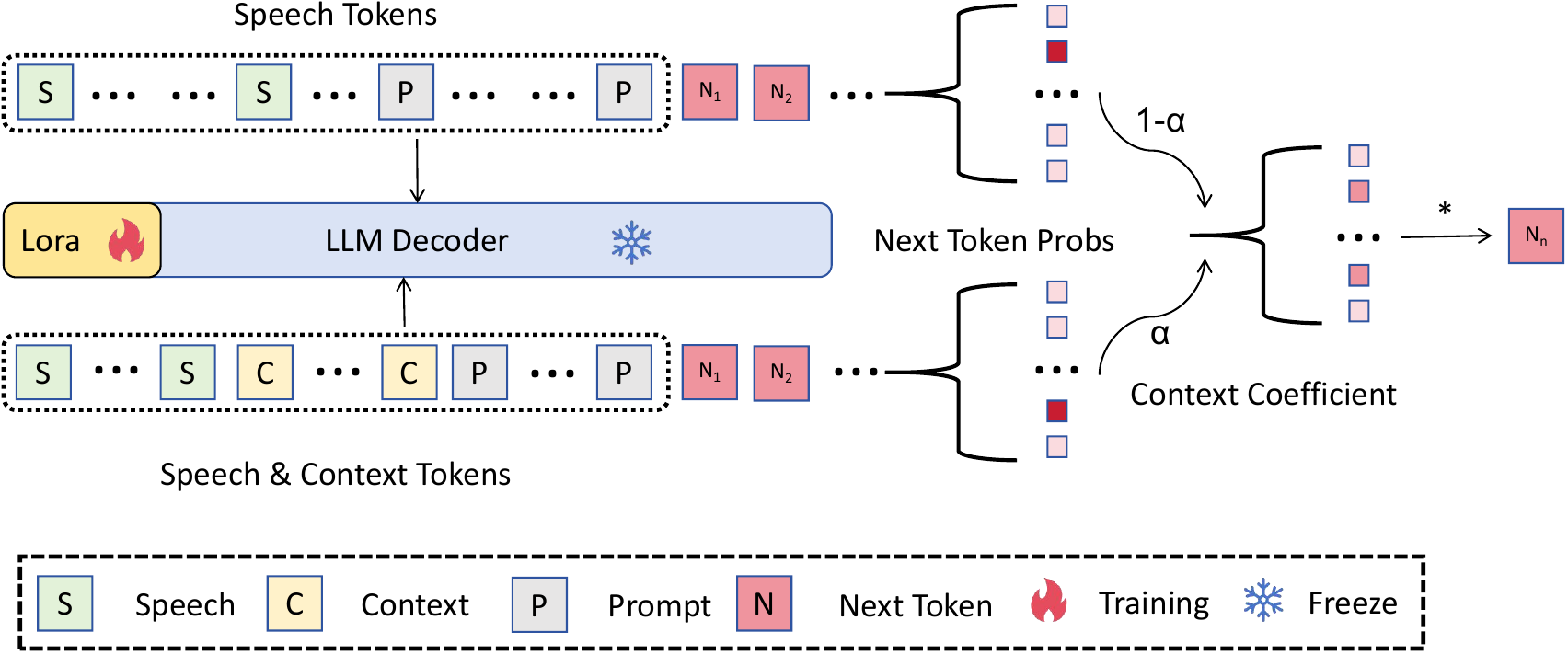}
\caption{Joint decoding method: By maintaining the decoding bundle with context and the decoding bundle without context, the two continuously merge the probability of next\_token during the decoding process, and force the fused decoding token to be the decoding result of the two bundles, so as to realize controllable context information injection.}
\label{Fig:joint_decode}
\end{figure*}

% 近年来，自动语音识别（Automatic Speech Recognition, ASR）领域正经历第三次范式变革：从传统系统～\cite {levinson1983introduction}，到端到端框架～\cite {graves2006connectionist,graves2012sequence,chan2016listen}，再到当前的语音大型语言模型（Speech LLM）范式～\cite {peng2024survey}，该领域正朝着具备人类水平的通用性与鲁棒性稳步迈进。其中，语音大型语言模型通过适配器实现语音编码器与大型语言模型的连接，在自动语音识别任务中取得了突破性进展，不仅斩获多项榜单的最优结果（SOTA），更展现出强劲的技术潜力～\cite {chen2024salm,yu2024connecting,fathullah2024prompting}。

 In recent years, the field of Automatic Speech Recognition (ASR) has been undergoing its third paradigm shift: from traditional systems \cite{levinson1983introduction,lee2002speaker,alharbi2021automatic}, to end-to-end frameworks \cite{graves2006connectionist, graves2012sequence, chan2016listen}, and further to the current Speech Large Language Model (Speech LLM) paradigm \cite{peng2024survey}, the field is steadily moving towards achieving human-level universality and robustness. Among these, Speech LLMs, which connect speech encoders to large language models via adapters, have made breakthrough progress in ASR tasks. They have not only achieved State-of-the-Art (SOTA) results across various benchmarks but also demonstrated strong technical potential \cite{chen2024salm, yu2024connecting, fathullah2024prompting}.
Regardless of the recognition paradigm, contextual automatic speech recognition (Contextual ASR)~\cite{aleksic2015bringing,hall2015composition} stands as a particularly critical issue in the field of ASR. Its core lies in leveraging contextual information to enhance the recognition accuracy of specific words—especially entity terms such as personal names and place names.

 Traditional ASR systems typically consist of a acoustic  model and a language model. So Google proposed the On-the-fly Rescore method~\cite{aleksic2015bringing,hall2015composition}, which adjusts the recognition probability of the language model for specific words using WFST during the decoding phase. In the era of deep learning, although the language model is no longer an essential component of the recognition system, it can still provide contextual bias to the system through methods like Shallow-Fusion~\cite{zhao2019shallow}; another approach is deep fusion, such as CLAS~\cite{pundak2018deep}, which embeds contextual information into the internal architecture of end-to-end models to achieve deep interaction between semantic information and acoustic features. In the era of Speech LLM, benefiting from the excellent contextual understanding ability of LLM, embedding contextual information into prompts has become the mainstream approach, and multiple studies have verified the effectiveness of this method ~\cite{chen2024salm,yang2024mala,shen2025retrieval,lakomkin2024end}. Lakomkin et al.~\cite{lakomkin2024end} investigated the recognition capability of Speech LLM when provided with contextual information such as audio titles and content descriptions, and explored the model's tolerance to contextual perturbations. Z. Chen et al.~\cite{chen2024salm} studied the recognition performance of Speech LLM in scenarios where entity words are given. G. Yang et al.~\cite{yang2024mala} examined the recognition capability of Speech LLM when utilizing multi-modal auxiliary information as context. P. Shen et al.~\cite{shen2025retrieval} proposed a method that combines Speech LLM with a Retrieval-Augmented Generation (RAG) system.

However, the current method of directly injecting contextual information through Prompts relies on an implicit Attention mechanism, making it difficult to precisely control the fusion intensity of contextual information. The need to emphasize contextual information varies across different scenarios, and over-emphasizing contextual information may interfere with the overall recognition accuracy. Therefore, a key question arises: \textbf{Can we explicitly control the degree of contextual information fusion based on Speech LLM to achieve better recognition results? Can we even suppress unwanted sensitive information?}

To address the above issues, this paper proposes a joint decoding method, which achieves control over the degree of contextual information fusion 
% by fusing two decoding bundles
. This method better controls the fusion of contextual information and achieves better results; meanwhile, it can effectively suppress the recognition of sensitive words under the premise of slightly reducing the recognition error rate. Finally, we find that even Speech LLM not trained with long contextual data can acquire long contextual capabilities through this method.

\section{Controllable context injection method based on Speech LLM}

\subsection{Contextual Speech Recognition in Speech LLM}

Let \( x \) denote the input speech information (acoustic feature sequence), \( c \) denote the given context information (such as entity words, scene descriptions, etc.), \( P \) denote the instruction information, and \( y = [y_1, y_2, \ldots, y_T] \) denote the prediction sequence generated by the model, where \( y_t \) represents the \( t \)-th token and \( T \) is the sequence length.
In the absence of context input, the decoding sequence is as follows:
\begin{equation}
P(y|x) = \prod_{t=1}^T P(y_t | y_{1:t-1}, x). 
\end{equation}

In the presence of context, the decoding sequence is as follows:
\begin{equation}
P(y|x,c) = \prod_{t=1}^T P(y_t | y_{1:t-1}, x, c). 
\end{equation}
\subsection{Joint Decode Method For Control Contextual Information}

   As shown in Fig~\ref{Fig:joint_decode} , Joint Decode achieves probabilistic fusion through weighted combination using a balance coefficient \( \alpha \), where the joint probability at each decoding step is calculated by linearly blending the context-dependent and context-independent probabilities. For the initial token \( y_0 \), the joint probability is
\begin{equation}
P_{\text{joint}} = \frac{\alpha}{1+\alpha} \cdot P(y_0|x,c) + \frac{1}{1+\alpha} \cdot P(y_0|x).
\end{equation}
 For subsequent tokens \( y_i \) (\( i \geq 1 \)), We force the jointly decoded token to be the common result of the two decoding bundle, it extends to
\begin{equation}
\resizebox{0.91\hsize}{!}{
$
P_{\text{joint}} = \frac{\alpha}{1+\alpha} \cdot P(y_i|y_{0:i-1},x,c) + \frac{1}{1+\alpha} \cdot P(y_i|y_{0:i-1},x)
$.
}
\end{equation}

% Hard joint decoding introduces a constraint mechanism via top-\( k \) tokens from context-independent predictions: first, extract the top-\( k \) tokens with the highest probabilities from the context-independent distribution \( P(y_i|y_{1:i-1},x) \), denoted as the set \( S_{\text{top-}k} = \{ y_i^{(1)}, y_i^{(2)}, \ldots, y_i^{(k)} \} \); then mask the context-dependent distribution \( P(y_i|y_{1:i-1},x,c) \) such that probabilities of tokens not in \( S_{\text{top-}k} \) are set to 0, i.e.,
% \begin{equation}
% \hat{P}{\text{masked}}(y_i|x,c) = \begin{cases} P(y_i|y{1:i-1},x,c) & \text{if } y_i \in S_{\text{top-}k}, \ 0 & \text{otherwise} \end{cases}
% \end{equation}
% ; finally, normalize the masked distribution to form a valid probability:
% \begin{equation}
% P_{\text{hard}} = \frac{\hat{P}{\text{masked}}(y_i|x,c)}{\sum{y_i' \in S_{\text{top-}k}} \hat{P}{\text{masked}}(y_i'|x,c)}
% \end{equation}
% , which is used directly for next-token decoding. Combined soft-hard decoding integrates the two strategies by performing weighted fusion on the masked context-dependent probabilities: after obtaining \( \hat{P}_{\text{masked}}(y_i|x,c) \) via the hard masking process, the joint probability is computed as
% \begin{equation}
% P{\text{combined}} = \frac{\alpha}{1+\alpha} \cdot \hat{P}{\text{masked}}(y_i|x,c) + \frac{1}{1+\alpha} \cdot P(y_i|y{1:i-1},x)
% \end{equation}
% , thus retaining both the constraint of top-\( k \) masking and the flexibility of weighted probability fusion.

\section{Experiment Setup}

\begin{table*}[!ht]
  \centering
  \caption{Basic Experiments on the Effects of Different Training Tasks on the Contextual Capabilities of Speech LLM}
  \label{tab:base_experiement}
  \renewcommand{\arraystretch}{1.1} % Adjust row spacing
  \newcolumntype{S}{>{\small}c} % Define small font centered column
  \begin{tabularx}{0.9\textwidth}{c|c|c|S|S|S}
    \toprule
    \multirow{2}{*}{\textbf{Training}} & 
    \multirow{2}{*}{\textbf{Context}} & 
    \multirow{2}{*}{\textbf{CER(\%)}} & 
    \multirow{2}{*}{\textbf{sub/insert/del}} & 
    \textbf{NE-Common} & 
    \textbf{NE-Rare} \\
    \cmidrule(lr){5-6}
    & & & & Precision/Recall/F1-Score & Precision/Recall/F1-Score \\
    \midrule
    \multirow{5}{*}{ASR} & None & 3.57 & 2.85/0.39/0.32 & 0.99/0.61/0.75 & 1.00/0.29/0.45 \\
    \cmidrule(lr){2-6}
    & NE-GT & \textbf{3.48 }& 2.47/0.39/0.59 & \textbf{1.00/0.81/0.89} & \textbf{1.00/0.65/0.79} \\
    \cmidrule(lr){2-6}
    & NE-Rare & 5.50 & 3.59/0.64/1.42 & 0.75/0.64/0.69 & 0.44/0.41/0.43 \\
    \cmidrule(lr){2-6}
    & NE-Common & 6.89 & 3.87/0.76/2.24 & 0.55/0.67/0.60 & 0.28/0.43/0.34 \\
    \midrule
    \multirow{5}{*}{ASR+NE-Short} & None & 3.38 & 2.80/0.34/0.24 & 0.98/0.54/0.70 & 1.00/0.23/0.37 \\
    \cmidrule(lr){2-6}
    & NE-GT & \textbf{2.56} & 2.02/0.32/0.22 & \textbf{1.00/0.87/0.93} & \textbf{1.00/0.79/0.88} \\
    \cmidrule(lr){2-6}
    & NE-Rare & 4.18 & 3.21/0.45/0.51 & 0.73/0.61/0.66 & 0.47/0.49/0.48 \\
    \cmidrule(lr){2-6}
    & NE-Common & 6.35 & 3.20/0.44/2.69 & 0.47/0.49/0.48 & 0.17/0.52/0.26 \\
    \midrule
    \multirow{5}{*}{ASR+NE-Long} & None & 3.28 & 2.75/0.33/0.24 & 0.98/0.55/0.70 & 0.99/0.23/0.38 \\
    \cmidrule(lr){2-6}
    & NE-GT & \textbf{2.49} & 2.00/0.31/0.17 & \textbf{1.00/0.87/0.93} & \textbf{1.00/0.76/0.87} \\
    \cmidrule(lr){2-6}
    & NE-Rare & 3.10 & 2.56/0.33/0.20 & 0.89/0.73/0.80 & 0.79/0.78/0.79 \\
    \cmidrule(lr){2-6}
    & NE-Common & 2.93 & 2.39/0.32/0.21 & 0.91/0.84/0.87 & 0.90/0.74/0.82 \\
    \bottomrule
  \end{tabularx}
\end{table*}

\subsection{DataSet Selection}
We take AISHELL-1~\cite{bu2017aishell} as our training dataset and entity words as the context information. During the training process, we utilize the LTP~\cite{che2020n} tool to extract entity words for each sentence. Then, we treat the Ground truth as NE-Short and combine the Ground Truth with any 5 to 50 entity words extracted from other sentences to form NE-Long. For the evaluation, we have the following entity word test sets: NE-GT, NE-Common, and NE-Rare. NE-GT is the ground truth extracted by our LTP tool, while the latter two are derived from the public data of \cite{shi2024seaco}, which include 400 common entity words and 233 rare entity words such as names of people and places. NE-Rare is included in NE-Common.

\subsection{Metric}

For general recognition performance, we use character error rate(CER) to measure it, and for contextual ability, we use entity word precision, recall,  and F1 value to measure it.

\subsection{Model Configuration}
Our Speech LLM follows the configuration of the current SOTA speech recognition model FireRed-LLM~\cite{xu2025fireredasr}, which consists of a Conformer~\cite{gulati2020conformer} encoder, a linear adapter, and an LLM decoder. We used the wenet tool to pre-train a Conformer AED model using 100,000 hours of our own data, and then used the Conformer encoder part of it. The Conformer encoder uses 12 attention heads, and the dimension of the middle layer of the feedforward network is 3584; the model finally outputs feature dimensions of 768, the convolution kernel size is 33, there are 16 layers in total. The implementation of the adapter is consistent with FireRed-LLM, and the folding rate is 2. LLM uses Qwen2.5-7B-Instruct~\cite{gulati2020conformer} and lora for fine-tuning. The lora configuration is consistent with FireRed-LLM, and the lora rank is 64. And there are a total of 441M training parameters.
\begin{table*}[!ht]
  \centering
  \caption{Experiments using Joint Decoding method to control contextual information injection in models trained with long/short context}
  \label{tab:experiment_results}
  \renewcommand{\arraystretch}{1.1} % Optimize row spacing
  \newcolumntype{S}{>{\small\centering\arraybackslash}X} % Small font centered column for metrics
  \begin{tabularx}{\textwidth}{c|c|c|c|S|S|S}
    \midrule
    
    \multirow{2}{*}{\textbf{Training}} &
    \multirow{2}{*}{\textbf{Context}} & 
    \multirow{2}{*}{\textbf{\makecell{Context \\Coefficient}}} & 
    \multirow{2}{*}{\textbf{CER (\%)}} & 
    \multirow{2}{*}{\textbf{sub/insert/del}} & 
    \textbf{NE-Common} & 
    \textbf{NE-Rare} \\
    \cmidrule(lr){6-7}
    & & & & & Precision/Recall/F1-Score & Precision/Recall/F1-Score \\
        \midrule
    \multirow{9}{*}{ASR+NE-Short} & None & 0.00 & 3.38 & 2.80/0.34/0.23 & 0.97/0.54/0.69 & 1.00/0.23/0.37 \\
    \cmidrule(lr){2-7}
    & \multirow{4}{*}{NE-Rare} & 0.30 & 3.42 & 2.80/0.37/0.24 & \textbf{0.89/0.62/0.73} & \textbf{0.76/0.47/0.58} \\
    \cmidrule(lr){3-7}
    & & 0.70 & \textbf{3.29} & 2.71/0.35/0.22 & \textbf{0.97/0.58/0.73} & 0.98/0.32/0.49 \\
    \cmidrule(lr){3-7}
    & & 1.00 & 4.18 & 3.21/0.45/0.51 & 0.72/0.60/0.66 & 0.47/0.49/0.48 \\
    \cmidrule(lr){2-7}
    & \multirow{4}{*}{NE-Common} & 0.30 & 3.36 & 2.73/0.36/0.26 & \textbf{0.92/0.66/0.77} & \textbf{0.84/0.47/0.60} \\
    \cmidrule(lr){3-7}
    & & 0.70 & \textbf{3.26} & 2.68/0.34/0.23 & 0.96/0.60/0.74 & 0.98/0.33/0.49 \\
    \cmidrule(lr){3-7}
    & & 1.00 & 6.35 & 3.20/0.44/2.69 & 0.47/0.49/0.48 & 0.17/0.52/0.26 \\
    \toprule
    
    \multirow{9}{*}{ASR+NE-Long} & None & 0.00 & 3.28 & 2.75/0.33/0.19 & 0.98/0.55/0.70 & 0.99/0.23/0.38 \\
    \cmidrule(lr){2-7}
    & \multirow{4}{*}{NE-Rare} & 0.30 & 3.06 & 2.55/0.32/0.18 & 0.98/0.64/0.77 & 0.99/0.46/0.63 \\
    \cmidrule(lr){3-7}
    & & 0.70 & \textbf{2.98} & 2.45/0.32/0.18 & \textbf{0.96/0.72/0.82} & \textbf{0.94/0.69/0.79} \\
    \cmidrule(lr){3-7}
    & & 1.00 & 3.10 & 2.56/0.33/0.20 & 0.89/0.73/0.80 & \textbf{0.79/0.78/0.79} \\
    \cmidrule(lr){2-7}
    & \multirow{4}{*}{NE-Common} & 0.30 & 3.20 & 2.49/0.31/0.18 & 0.98/0.67/0.80 & 0.99/0.43/0.60 \\
    \cmidrule(lr){3-7}
    & & 0.70 & \textbf{2.88} & 2.37/0.31/0.19 & \textbf{0.96/0.80/0.87} & 0.98/0.64/0.77 \\
    \cmidrule(lr){3-7}
    & & 1.00 & 2.93 & 2.39/0.32/0.21 & \textbf{0.91/0.83/0.87} & \textbf{0.90/0.74/0.81} \\

    \bottomrule
  \end{tabularx}
\end{table*}
\subsection{Training Strategy}
Our Speech LLM will undergo a two-stage training process. In the first stage, the encoder and adapter are trained using ASR tasks. The second stage then involves training all trainable components through NE-Short/NE-Long contextual ASR tasks.  For the ASR task, the prompt is 
\begin{CJK}{UTF8}{gbsn} \textbf{请识别语音并转写为文字: \textless $\vert $speech$\vert$ \textgreater } \end{CJK} ; for the contextual ASR task, the prompt is \begin{CJK}{UTF8}{gbsn} \textbf{请识别语音并转写为文字,下面的热词可能会提供帮助，... : \textless $\vert $speech$\vert$ \textgreater } \end{CJK} .

% 模型
    % Our Model
    % OpenSource: FireredASR
% 数据:
    % aishell-1
    % 热词数据：
    % 训练数据：
        % gt
        % gt_20~200
    % 评估数据：
        % 自有的
        % gt
        % gt_200
        % gt_400
        % gt_xxx
        % 公开的
        % Test-Aishell1-NE-R1 223
        % Test-Aishell1-NE 400
% 指标
    % CER
    % F1/Recall/Precise
    % NE-CER / NE-FNR

% 方法有有效性：更优秀的控制
% 模型：
    % 自己训练的 gt_20-200
% gt : 联合解码
% Test-Aishell1-NE-R1 223  联合解码
% Test-Aishell1-NE 400  联合解码
%
% 抑制实验：
% 模型：
    % 自己训练的 gt_20-200
% gt : 联合解码
% Test-Aishell1-NE-R1 223  联合解码
% Test-Aishell1-NE 400  联合解码

% 未经过训练的模型可以具备热词能力吗
% 模型：
    % 自己训练：
    % FireRedASR
% 热词：
   % gt : 联合解码
    % Test-Aishell1-NE-R1 223  联合解码
    % Test-Aishell1-NE 400  联合解码

% 训练热词对热词能力的影响
% 模型：
    % 自己训练的 gt_20-200
    % 自己训练的 gt
% gt : 联合解码
% Test-Aishell1-NE-R1 223  联合解码
% Test-Aishell1-NE 400  联合解码
% 热词规模 xxx
% 、

\section{Result and Discusion}

% As the table~\ref{tab:base_experiement} shows, we trained three basic models using ASR tasks, ASR tasks plus  NE-Short context tasks, and ASR tasks plus  NE-Long cotext tasks. We tested the recognition results of the three basic models in four situations: no context, NE-GT,  NE-Rare and NE-Common, in order to show the impact of training on the model's contextual capabilities. 
% 从表格中我们可以观察到，在给定NE-GT的情况下，即便是没有经过上下训练，Speech LLM也具备的一定的上下文能力，比如只经过ASR训练的模型，在给定NE-GT的上下文信息之后，CER从3.57下降到了3.48。不过经过了上下文训练确实能够增强Speech LLM的上下文能力，分别经过NE-Short和NE-Long后的Speech LLM在给定NE-GT之后，错误率分别从3.38下降到了2.56和从3.28下降到了2.49。然而很遗憾的是，当给定NE-Rare/Common时，无论是没有经过上下文训练，还是经过了短上下文训练，当给定过长上下文时，识别结果都出现了下降，主要是由于替换错误和插入错误的剧烈增加。这意味着长上下文训练依然是有必要的。经过NE-Long训练过的模型，在给定NE-Rare和NE-Common后，识别率分别从3.28下降到了3.10和2.93.
% \subsection{Basic Experiments on the Effects of Different Training Tasks on the Contextual Capabilities of Speech LLM}
As shown in Table~\ref{tab:base_experiement}, we trained three basic models using ASR tasks alone, ASR tasks combined with NE-Short context tasks, and ASR tasks combined with NE-Long context tasks. We tested the recognition results of these three basic models under four scenarios: no context, NE-GT, NE-Rare, and NE-Common, aiming to demonstrate the impact of training on the model's contextual capabilities. It can be observed from the table that when NE-GT is provided, the Speech LLM already possesses a certain degree of contextual capability even without undergoing contextual training. For example, the model trained solely with ASR sees its CER decrease from 3.57 to 3.48 after being given NE-GT context information. However, contextual training does enhance the Speech LLM's contextual capabilities. The Speech LLMs trained with NE-Short and NE-Long respectively show their error rates drop from 3.38 to 2.56 and from 3.28 to 2.49 when provided with NE-GT. Unfortunately, when given NE-Rare/Common, whether the model has not undergone contextual training or has only received short-context training, the recognition results decline when provided with overly long contexts. This is mainly due to a sharp increase in substitution errors and insertion errors. This indicates that long-context training is still necessary. The model trained with NE-Long sees its CER decrease from 3.28 to 3.10 and 2.93 when provided with NE-Rare and NE-Common respectively. 

% 我们训练将联合解码方法运用于控制上下文信息注入时，取得了更好的成绩，如表~\ref{tab:experiment_results}所显示。可以观察到伴随着融合系数增加，实体词的准确率下降，召回率上升，F1值先升高后下降，CER也同样先升高后下降。最优融合系数为0.7，在此时长/短上下文训练训练的模型在给定NE-Rare和NE-Common的情况下均取得了最好的结果，CER分别为2.98/2.88和3.29/2.26。令人惊讶的是即使只经过短上下文训练的模型也具备了长上下文的能力，当然仍然比不过经过长上下文训练的模型。

\begin{table}[htbp]
  \centering
  \caption{Experiments using a joint decoding method to suppress sensitive words; the basic model is trained with long context.}
  \renewcommand{\arraystretch}{1.1} % Increase row height for readability
  \begin{tabular}{c|c|c|c}
    \hline
    \multirow{2}{*}{\textbf{Context}} & \multirow{2}{*}{\makecell{\textbf{Context} \\ \textbf{Coefficient}}} & \multirow{2}{*}{\textbf{CER}} & \textbf{NE-Rare} \\
    \cline{4-4}
    & & & \makecell{Precision/Recall\\/F1-Score} \\
    \hline
    None & 0.00 & 3.28 & 0.99/0.23/0.38 \\
    \hline
    \multirow{4}{*}{NE-Rare} & -0.10 & 3.36 & 1.00/0.17/0.30 \\
    \cline{2-4}
    & -0.20 & 3.42 & 1.00/0.15/0.26 \\
    \cline{2-4}
    & -0.50 & 3.90 & 1.00/0.07/0.13 \\
    \cline{2-4}
    & -1.00 & 4.75 & 1.00/0.01/0.03 \\
    \hline
  \end{tabular}
  \label{tab:experiment_results_sorted}
\end{table}

When applying the joint decoding method to control the context information injection, we achieve better results, as shown in Table~\ref{tab:experiment_results}. It can be observed that as the fusion coefficient increases, the precision of entity words decreases, the recall increases, the F1 score first increases and then decreases, and the CER also first increases and then decreases. The optimal fusion coefficient is 0.7, at which the models trained with both long and short contexts achieve the best results given NE-Rare and NE-Common, with CERs of 2.98/2.88 and 3.29/2.26, respectively. Surprisingly, even models trained with only short contexts demonstrate the ability to handle long contexts, though they still fall short of models trained with long contexts. It can be observed that the reason why joint decoding can achieve better recognition results is that the highest Precision is achieved, which greatly reduces the substitution error and slightly reduces other errors.

% \subsection{Basic Experiments on the Effects of Different Training Tasks on the Contextual Capabilities of Speech LLM}
% 如表~\ref{tab:experiment_results_sorted}所显示，我们在经过长上下文训练的模型的基础上，将NE-Rare作为敏感词进行抑制的实验，实验显示我们的方法可以有效的抑制敏感词，当给定系数为-1时，敏感词召回率下降为0.01
As shown in Table~\ref{tab:experiment_results_sorted}, we experimented with NE-Rare as a sensitive word for suppression based on the model trained with long context. The experiment showed that our method can effectively suppress sensitive words. When the given coefficient is -1, the recall rate of sensitive words drops to 0.01.

\section{Conclusion And Future Work}

% 当前的使用Speech LLM的上下文能力和在Prompt中注入上下文信息进行Context ASR的方法无法显示的控制上下文信息的融入程度，针对这个问题我们提出了联合解码方法，实验证明在我们的方法下可以通过控制融合程度更好的注入上下文信息并且获得更好的识别结果，甚至可以有效的抑制敏感词出现。

The current method of using the contextual capabilities of Speech LLM and injecting contextual information into prompts for Context ASR cannot explicitly control the degree of integration of contextual information. To address this problem, we proposed a joint decoding method. Experiments have shown that with our method, we can better inject contextual information and obtain better recognition results by controlling the degree of integration, and even effectively suppress the appearance of sensitive words. 

However, maintaining two decoding beams simultaneously increases the decoding cost. In the future, we will explore more efficient mechanisms to achieve explicit control over contextual information.

\bibliographystyle{IEEEbib}
\bibliography{main}

\end{document}